\documentstyle[12pt]{article} 

\newcommand{\EQ}{\begin{equation}}
\newcommand{\EN}{\end{equation}}
\newcommand{\bea}{\begin{eqnarray}} 
\newcommand{\ena}{\end{eqnarray}}
\newcommand{\vs}[1]{\vspace{#1 mm}}

\renewcommand{\a}{\alpha}

\renewcommand{\d}{\delta}
\newcommand{\e}{\epsilon}

\def\bbox{{\,\lower0.9pt\vbox{\hrule \hbox{\vrule height 0.2 cm
\hskip 0.2 cm \vrule height 0.2 cm}\hrule}\,}}
\newcommand{\dsl}{\pa \kern-0.5em /}

\newcommand{\pa}{\partial}

\renewcommand{\k}{\kappa}

\newcommand{\nn}{\nonumber\\}
\newcommand{\p}[1]{(\ref{#1})}

\begin{document}

\topmargin 0pt
\oddsidemargin 0mm
\renewcommand{\thefootnote}{\fnsymbol{footnote}}
\begin{titlepage}

\setcounter{page}{0}
\begin{flushright}
TUW-01-04\\
\end{flushright}

\vs{10}
\begin{center}

{\Large\bf D-branes in B Fields}

\vs{15}

{\large 
Jian-Ge Zhou\footnote{e-mail address: jgzhou@hep.itp.tuwien.ac.at}}

\vs{10}
{\em Institut f\"ur Theoretische Physik, Technische Unversit\"at Wien,\\
Wiedner Hauptstrasse 8-10/136, A-1040 Wien, Austria} \\
\end{center}

\vs{15}
\centerline{{\bf{Abstract}}}
\vs{5}
The RR Page charges for the D2-, D4-, D6-brane in B fields
are constructed explicitly from the equations of motion and the 
nonvanishing
(modified) Bianchi identities by exploiting their properties  
 --- conserved and localized.
It is found that the RR Page charges are independent of
the backgound B fields, which provides further evidence that
the RR Page charge should be quantized. In our construction,
it is highly nontrivial that the terms like $B\wedge B\wedge B$,
$B\wedge B\wedge F$, $B\wedge F\wedge F$ from different sources
are exactly cancelled with each other.
\end{titlepage}
\newpage

\renewcommand{\thefootnote}{\arabic{footnote}}
\setcounter{footnote}{0}

{\Large\bf 1. Introduction}
\vs{5}

In IIA string theory, a D2p-brane is coupled to the RR gauge 
fields~\cite{jp}, the coupling of the RR gauge fields to the pullback
of the spacetime B field and U(1) gauge field strength F on  
D2p-brane seems to demand the integral
$\int (B+F)$ to be quantized. However,  in~\cite{bds},
Bachas, Douglas and Schweigert found that in WZW model the RR D0-brane
charge on the stable spherical D2-brane takes irrational value
due to the integral $\int B$, and suggested only $\int F$
can be interpreted as properly defined RR D0-brane charge.

Up to now, three different approaches have been proposed to 
resolve
this puzzle. In~\cite{wt}, Taylor argued that the D0-brane
charge arising from the integral over the D2-brane of the
pullback of the B field is cancelled by the bulk contributions,
but in his calculation it was implicitly assumed that the
gauge field $C^{(1)}$ is constant. Also his analysis is hard to 
be generalized to D2p-branes ($p > 1$)\footnote{Here we mean that
when $p = 2,3$, we have D4-, D6-brane.}. In~\cite{amm},
Alekseev, Mironov and Morozov observed that if one rewrites the
WZ action in terms of the gauge invariant field strength 
$G^{(2p+2)} = dC^{(2p+1)} - C^{(2p-1)}\wedge H$,
the resulting RR brane charges are independent of B fields.
However, in their approach when the RR gauge fields $C^{(2p+1)}$
are not constant they have to deal with some
nonlocal transformation, and treat RR gauge fields $C^{(2p+1)}$
as independent fields even though in IIA string theory
the RR gauge field $C^{(2p+1)}$ and $C^{(7-2p)}$ are dual to each
other. In~\cite{dm}, Marolf suggested that due to the presence
of Chern-Simons term in IIA supergravity, there are 
three notions of charge: 1) `brane source charge'
is gauge invariant and localized, but not conserved; 2) `Maxwell
charge' is conserved and gauge invariant, but not localized; 
3) `Page charge' is conserved, localized and invariant under
small gauge transformation. Since brane source charge
and Page charge were defined separately in~\cite{dm},
it is unclear how to relate one to other which is crucial 
to the questions whether it is possible for us to consistently 
define the general RR Page charges for a D2p-brane from
brane source charges and whether Page charges are 
independent of B fields or not\footnote{In general, the definition
of RR Page charge should be compatible with T-duality, and 
quantization of the Page charge for a D2p-brane follows from
T-duality. T-duality implies Page charge quantization in
the systems with sufficient translational symmetry~\cite{dm}.}. 

In the present paper, 
we study how to consistently construct 
B-independent RR charges of D2p-branes 
with B fields in our new frame.
Since D-branes are sources of RR gauge fields, we consider
IIA supergravity plus D-brane sources, from which we derive the equations
of motion and the nonvanishing (modified) Bianchi identities that
define the duals of brane source currents for D2p-branes.
Exploiting the relation $dA_{p}\wedge B_{q} = d(A_{p}\wedge B_{q}) - 
(-1)^{p}A_{p}\wedge dB_{q}$, we insert the equations for the
duals of brane source currents for D(2p+2n)-branes ($n > 1$) into
that for the D2p-brane iteratively. The new feature is that the resulting
equations can be recast into the form whose left sides of equations
are exterior derivative and the right sides are localized objects,
thus the right side localized objects can be identified
as Page charges because of their conservation and locality.
The results also show that the Page charges for D2p-branes can
be consistently defined from the brane source charges. Inserting
the brane source charges into the expression of the Page charges,
we find that all the Page charges are independent of the background
B fields. In our explicit construction, it is highly nontrivial
that the B-dependent terms like $B\wedge B\wedge B$,
$B\wedge B\wedge F$, $B\wedge F\wedge F$ from different sources
are exactly cancelled with each other.

The layout of the paper is as follows. In section 2, we
consider a D2-brane in B-fields in order to illustrate our method. 
The brane source currents for D2-, D0-branes are derived from the equations
of motion.  The explicit expression
for RR Page charges is obtained by exploiting their properties  ---
conserved and localized, the result is consistent with those in ~\cite{wt} and
~\cite{amm}. In section 3, we study a D4-brane in B fields,
which is bound state of D4/D2/D0-branes. With additional
D4-brane source defined by the nonvanishing of modified
Bianchi identity, the RR Page charges for
D2-, D0-branes living on the D4-brane are constructed by combining 
the equations of motion and
the nonvanishing of the modified Bianchi identity.
Late we discuss a D6-brane in B fields, where we find that it is
highly nontrivial that the RR Page charges are independent of 
B fields. In section 4, we present our summary and discussion.

\vs{5}
{\Large\bf 2. D2-brane in B fields}
\vs{5}

The IIA supergravity in ten dimensions can be obtained from
$N = 1$, $D = 11$ supergravity by dimensional reduction,
the action for IIA supergravity is
\bea
S_{IIA}&=&\frac{1}{2\k_{10}^{2}}\int d^{10}x\sqrt{-G} 
\Big\{e^{-2\phi}(R + 4\pa_{\mu}\phi\pa^{\mu}\phi - \frac{1}{2}|H|^{2})\nn
&&
-\frac{1}{2}(|F^{(2)}|^{2} + |\tilde {F}^{(4)}|^{2})\Big\}
-\frac{1}{4\k_{10}^{2}}\int B\wedge F^{(4)}\wedge F^{(4)}
\ena
where $\k_{10}^{2}$ is the gravitational coupling in
ten dimensions, and the field strengths $H,F^{(2)}, F^{(4)}$, and 
$\tilde {F}^{(4)}$ are defined as 
\EQ
H = dB,~~ F^{(2)} = dC^{(1)},~~ F^{(4)} = dC^{(3)},~~ \tilde {F}^{(4)}
= dC^{(3)} - C^{(1)}\wedge H
\EN

The above action can be rewritten as
\bea
S_{IIA}&=&\int d^{10}x\sqrt{-G} 
e^{-2\phi}(R + 4\pa_{\mu}\phi\pa^{\mu}\phi)\nn
&&
+ \frac{1}{2}\int \Big\{-e^{-2\phi} dB\wedge *dB + dC^{(1)}\wedge *dC^{(1)}\nn
&&
+ (dC^{(3)} - C^{(1)}\wedge dB)\wedge (dC^{(3)} - C^{(1)}\wedge dB)\nn
&&
- B\wedge dC^{(3)} \wedge dC^{(3)}\Big\}
\label{2a}
\ena
where for simplicity we have chosen $2\k_{10}^{2} = 1$, that is,
the kinetic term is canonical.

Consider a D2p-brane in the background of B fields, the D2p-brane
world-volume action is
\EQ
S_{D-brane} = S_{BI} + S_{WZ}
\EN
The Born-Infeld part of the action is 
\EQ
S_{BI} = -T_{2p}\int e^{-\phi}\sqrt{-det(G_{ab} + B_{ab} + F_{ab})}
\EN
where $G_{ab}$, $\phi$, $B_{ab}$ are the pullback of 
spacetime metric, dilaton and Neveu-Schwarz two form B to the D2p-brane
world-volume, $F_{ab}$ is the field strength of U(1) gauge field
living on the brane.

The Wess-Zumino (WZ) terms couple the brane to the spacetime RR gauge
field through
\EQ
S_{WZ} = \int ( \sum\limits_{i} C^{(i)})\wedge e^{B+F}
\EN
and in the D2-brane case, it is reduced to 
\EQ
S_{WZ}^{D2}= \int\Big\{C^{(3)} + C^{(1)}\wedge (B+F)\Big\}
\EN
which means there are D0-branes living on D2-brane.

Since D-branes are sources of RR gauge field, we consider
IIA supergravity + the coupled D2-brane, and see how 
D2-, D0-branes induce $C^{(3)}$, $C^{(1)}$ gauge field,
which can be described by the equations of motion.
As brane source charge arises from varying the brane action
with respect to the gauge field, from the total action 
$S_{T} = S_{IIA} + S_{D2-brane}$ we get the equations of
motion for $C^{(1)}$ and $C^{(3)}$\footnote{Since we choose different
sign for the WZ term, there is minus sign difference between our \p{eqm}
and that in~\cite{dm}, similarly for the definition of 
the nonvanishing Bianchi identities.}
\bea
d*F^{(2)} - H\wedge *\tilde{F}^{(4)} &=& -*j^{bs}_{D0}\nn
d*\tilde{F}^{(4)} - H\wedge\tilde{F}^{(4)}&=&-*j^{bs}_{D2}
\label{eqm}
\ena
where we have replaced the second term $H\wedge F^{(4)}$ in the
left side of the second equation by $H\wedge\tilde{F}^{(4)}$
due to the fact $H\wedge H = 0$. The brane source currents 
for D2-, D0-branes are given by
\EQ
(j_{D2}^{bs})^{\mu\nu\rho}(x) = \int dX^{\mu}(\xi)\wedge dX^{\nu}(\xi)\wedge
dX^{\rho}(\xi)\d^{10}(x-X(\xi))\nn
\EN
\EQ
(j_{D0}^{bs})^{\mu}(x) = \frac{1}{2}\int d^{3}\xi \e^{abc}\pa_{a}X^{\nu}(\xi)
[B_{\nu\rho}(X(\xi))\pa_{b}X^{\nu}(\xi)\pa_{c}X^{\rho}(\xi) + F_{ab}(\xi)]
\d^{10}(x-X(\xi))
\EN
where $x^{\mu}$ are ten-dimensional coordinates, $X^{\nu}(\xi)$ are 
the embedding coordinates of D2-brane in ten dimensions, and
$\xi^{a}$ are three-dimensional brane worldvolume coordinates.
The brane source charge for D0-brane is 
\EQ
Q_{D0}^{bs} = \int *j_{D0}^{bs} = \int\limits_{V_2} (B + F)
\EN
where ${V_2}$ is two-dimensional space volume of D2-brane.

Since we only consider D2-, D0-brane without other brane sources, 
we have $dF^{(2)} = 0$,
$dF^{(4)} = 0$, $dH = 0$, $H\wedge H = 0$. Under these conditions,
we see $d*j_{D2}^{bs} = 0$, $d*j_{D0}^{bs} \not= 0$, that is,
$Q_{D2}^{bs}$ is conserved while $Q_{D0}^{bs}$ nonconserved, however,
when $H = dB = 0$, $Q_{D0}^{bs}$ is still conserved.

By rewriting the second term $H\wedge *\tilde{F}^{(4)}$ in the first
equation in \p{eqm} as $d(B\wedge *\tilde{F}^{(4)})
- B\wedge d*\tilde{F}^{(4)}$ and then inserting the second equation of  
\p{eqm} into $B\wedge d*\tilde{F}^{(4)}$, we arrive at
\EQ
d\Big\{-*F^{(2)} + B\wedge *\tilde{F}^{(4)} - \frac{1}{2}B\wedge 
B\wedge\tilde{F}^{(4)}\Big\} =
*j_{D0}^{bs} - B\wedge *j_{D2}^{bs}
\label{d0pc}
\EN
which shows that we can indeed recast two equations in \p{eqm}
into the new equation whose left side is written as
exterior derivative and the right side as localized object.
In the above derivation, we have assumed that the D2-brane stayes
in the region where the relation $H = dB$ holds.
According to the properties of the RR Page current, Eq.\p{d0pc} 
indicates that we can define consistently
\EQ
*j_{D0}^{Page} = *j_{D0}^{bs} - B\wedge *j_{D2}^{bs}
\label{d0page}
\EN
and $j_{D0}^{Page}$ is indeed conserved and localized.
The RR Page charge takes
\EQ
Q_{D0}^{Page} = \int *j_{D0}^{Page} = \int\limits_{V_{2}}F
\EN
which is consistent with~\cite{wt} and~\cite{amm}. In~\cite{bds}
and~\cite{dm}, it was argued that the RR Page charge for D0-branes in
spherical D2-brane defined by $\int\limits_{S_{2}}F$ should be
quantized. In the context of WZW model, it has been shown that the 
integral $\int\limits_{S_{2}}F$ is quantized indeed~\cite{kkz}. Other related
discussion on this topic can also be found in~\cite{cs},~\cite{fs}
and~\cite{kz}.

If we do T-duality along other extra four dimensions,
the D2/D0 bound state turns into D6/D4 one, and
Eq.\p{d0page} becomes 
\EQ
*j_{D4}^{Page} = *j_{D4}^{bs} - B\wedge *j_{D6}^{bs}
\label{tpage}
\EN
which means we define the RR Page current for D4-branes living
on the D6-brane in the way of compatible with T-duality\footnote{Here
we should mention that \p{tpage} is obtained just by the use
of T-duality as suggested by Marolf.}.

\vs{5}
{\Large\bf 3. D4-, D6-brane in B fields}
\vs{5}

At first. we consider a D4-brane in B fields, which
is D4/D2/D0 bound state. The WZ term for D4-brane 
in the background of B fields is
\EQ
S_{WZ}^{D4} = \int \Big\{C^{(5)} + C^{(3)}\wedge (B+F) + \frac{1}{2}C^{(1)}
\wedge (B+F)\wedge (B+F) \Big\}
\label{d4wz}
\EN
Besides the equations of motion \p{eqm}, there is the nonvanishing
of modified Bianchi identity which describes the hodge dual of
the D4-brane source
current~\cite{dm}
\EQ
d\tilde{F}^{(4)} + F^{(2)}\wedge H = -*j^{bs}_{D4}
\label{mbi}
\EN
If we rewrite IIA supergravity in terms of the field $C^{(5)}$
dual to $C^{(3)}$, the modified Bianchi identity
for $C^{(3)}$ becomes an equation of motion for $C^{(5)}$, and
the brane source current can be derived by variation
of the WZ term for D4-brane with respect to $C^{(5)}$. From Eq.\p{d4wz},
we know the brane
source charges for D2-, D0-branes in the D4-brane are
\bea
Q_{D2}^{bs} = \int *j_{D2}^{bs} &=& \int (B + F)\nn
Q_{D0}^{bs} = \int *j_{D0}^{bs} &=& \int\frac{1}{2}(B + F)\wedge(B + F)
\ena
In the presence of D4-brane, $dF^{(4)}$ is no longer equal to
zero, then $Q_{D2}^{bs}$ is nonconserved. Without D6-brane,
we have $dF^{(2)} = 0$, so $Q_{D4}^{bs}$ is still conserved.
Combining Eq.\p{eqm} and Eq.\p{mbi}, we have
\EQ
d\Big\{-*\tilde{F}^{(4)} + B\wedge\tilde{F}^{(4)} + \frac{1}{2}B\wedge B\wedge
F^{(2)}\Big\} = *j_{D2}^{bs} - B\wedge *j_{D4}^{bs}
\label{d4d2}
\EN
\EQ
d\Big\{-*F^{(2)} + B\wedge *\tilde{F}^{(4)} - \frac{1}{2}B\wedge B\wedge
\tilde{F}^{(4)} - \frac{1}{6}B\wedge B\wedge B\wedge F^{(2)}\Big\}
= *j_{D0}^{bs} - B\wedge *j_{D2}^{bs} + \frac{1}{2}B\wedge B\wedge 
*j_{D4}^{bs}
\label{d4d0}
\EN
where we use the similar way as in the derivation
of Eqs.\p{d0pc} to get Eqs.\p{d4d2}. To derive \p{d4d0}, 
we first rewrite the 
second term $H\wedge *\tilde{F}^{(4)}$ in the first equation in \p{eqm} 
as $d(B\wedge *\tilde{F}^{(4)}) 
- B\wedge d*\tilde{F}^{(4)}$ and insert the second equation of 
\p{eqm} into $B\wedge d*\tilde{F}^{(4)}$, then besides the exterior derivative
and localized terms we have the term $B\wedge H\wedge\tilde{F}^{(4)}$
which can rewrite as $d(\frac{1}{2}B\wedge B\wedge
\tilde{F}^{(4)}) - \frac{1}{2}B\wedge B\wedge d\tilde{F}^{(4)}$.
Inserting the nonvanishing modified Bianchi identity \p{mbi} into the term
$\frac{1}{2}B\wedge B\wedge d\tilde{F}^{(4)}$ and exploiting 
$dF^{(2)} = 0$, we get Eq.\p{d4d0}.
Eqs.\p{d4d2} and \p{d4d0} show that we can consistently define
the hodge duals of Page currents
\bea
*j_{D2}^{Page} &=& *j_{D2}^{bs} - B\wedge *j_{D4}^{bs}\nn 
*j_{D0}^{Page} &=& *j_{D0}^{bs} - B\wedge *j_{D2}^{bs} + \frac{1}{2}
B\wedge B\wedge *j_{D4}^{bs}
\label{d4d0c}
\ena
from the brane source currents, the resulting $j_{D2}^{Page}$ and 
$j_{D0}^{Page}$ are conserved and 
localized on D4-brane, and the RR Page charges are 
\bea
Q_{D2}^{Page} &=& \int *j_{D2}^{Page} = \int F\nn
Q_{D0}^{Page} &=& \int *j_{D0}^{Page}
= \int\Big[\frac{1}{2}(B+F)\wedge (B+F) - B\wedge (B+F) + 
\frac{1}{2}B\wedge B\Big]\nn
&=& \int\frac{1}{2}F\wedge F
\label{d4d0p}
\ena
where we have seen the B-dependent terms 
$B\wedge B$, $B\wedge F$ have been cancelled with
each other.

Now we turn to a D6-brane in B fields, the WZ terms for D6-brane is 
\bea
S_{WZ}^{D6} &=& \int \Big\{C^{(7)} + C^{(5)}\wedge (B+F) + \frac{1}{2}C^{(3)}
\wedge (B+F)\wedge (B+F)\nn
&&
+ \frac{1}{6}C^{(1)}
\wedge (B+F)\wedge (B+F)\wedge (B+F)\Big\}
\label{wzd6}
\ena
which describes D6/D4/D2/D0 bound state. In the presence of D6-brane, 
besides the equations of motion \p{eqm} and the modified Bianchi
identity \p{mbi}, we have another nonvanishing Bianchi identity\footnote{In
IIA supergravity, the RR gauge field $C^{(2p+1)}$ is dual to $C^{(7-2p)}$.
Because of $**A_{2p} = -A_{2p}$ in 10D Minkowski spacetime, there are two 
ways to performing the electromagnetic duality. For example, consider
D6-brane RR gauge field $C^{(7)}$ (for illustration, we assume B fields 
vanish), we can define either ~~~$*dC^{(7)} = dC^{(1)}$, i.e., 
$*dC^{(1)} = -dC^{(7)}$, or ~~~$*dC^{(1)} = dC^{(7)}$, i.e., 
$*dC^{(7)} = -dC^{(1)}$. In case 1, the D6-brane source current is 
described by $dF^{(2)} = *j^{bs}_{D6}$, and in case 2, it is 
$dF^{(2)} = -*j^{bs}_{D6}$. The WZ term does not contain any
information about which definition one should use~\cite{dl}. In order
to make the definition for RR Page current compatible with T-duality,
we have to choose $dF^{(2)} = *j^{bs}_{D6}$.}
\EQ
dF^{(2)} = *j^{bs}_{D6}
\label{bi}
\EN
Combining two nonvanishing Bianchi identities \p{mbi} and \p{bi}, we have
\EQ
-d\Big(\tilde{F}^{(4)} + B\wedge F^{(2)}\Big) = 
*j^{bs}_{D4} - B\wedge *j^{bs}_{D6}
\label{my}
\EN
which shows that we can define the hodge dual of RR Page current for 
D4-branes living
in D6-brane as\footnote{If we choose $dF^{(2)} = -*j^{bs}_{D6}$,
then we would have $*j_{D4}^{Page} = *j_{D4}^{bs} + B\wedge *j_{D6}^{bs}$,
which is incompatible with T-duality.}
\EQ
*j_{D4}^{Page} = *j_{D4}^{bs} - B\wedge *j_{D6}^{bs}
\label{d4d6}
\EN
which is conserved and localized.
Here we should point out that Eq.\p{d4d6} is derived from the
nonvanishing modified Bianchi identity \p{mbi} for D4-brane 
and the other nonvanishing Bianchi identity \p{bi} for D6-brane.
On the other hand, Eq.\p{tpage} is obtained from T-duality, which
means the ambiguity in the definition of $C^{(7)}$ can be fixed by
T-duality.

In~\cite{dm}, the hodge dual of RR Page current for D4-brane was defined as
\footnote{In~\cite{dm}, the brane source charge and Page charge
were defined separately, they did not study how to define
Page charge based on the definition for brane source charge, and
\p{don} was given just by definition, but here \p{d4d6} is derived
from the consistent consideration.}
\EQ
-d(\tilde{F}^{(4)} +  C^{(1)}\wedge H) = *j^{Page}_{D4}
\label{don}
\EN
Comparing the Eq.\p{my} with Eq.\p{don}, we find there is
a difference in the definition of the RR Page current, which
can be interpreted as gauge dependence of the RR Page charge
associated with choosing a surface in $N = 1$ eleven-dimensional
supergravity that projects onto the chosen surface in IIA
supergravity.

By making use of the equations of motion and two nonvanishing (modified)
Bianchi identities, in the way similar to deriving Eqs.\p{d4d2} and \p{d4d0}
we get the relations
\EQ
d\Big(-*\tilde{F}^{(4)} + B\wedge \tilde{F}^{(4)} + \frac{1}{2}B\wedge B\wedge
F^{(2)}\Big) = *j_{D2}^{bs} - B\wedge *j_{D4}^{bs} + \frac{1}{2}B\wedge B\wedge
*j_{D6}^{bs}
\label{rela1}
\EN
\bea
d\Big(-*F^{(2)} + B\wedge *\tilde{F}^{(4)} - \frac{1}{2}B\wedge B\wedge
\tilde{F}^{(4)} - \frac{1}{6}B\wedge B\wedge B\wedge F^{(2)}\Big)\nn
= *j_{D0}^{bs} - B\wedge *j_{D2}^{bs} + \frac{1}{2}B\wedge B\wedge
*j_{D4}^{bs} - \frac{1}{6}B\wedge B\wedge B\wedge *j_{D6}^{bs}
\label{rela2}
\ena
Eqs.\p{rela1} and \p{rela2} indicate that the hodge dual
of the RR Page currents for 
D2-, D0-branes can be defined as
\EQ
*j_{D2}^{Page} = *j_{D2}^{bs} - B\wedge *j_{D4}^{bs} + 
\frac{1}{2}B\wedge B\wedge *j_{D6}^{bs}
\label{d2d6}
\EN
\bea
*j_{D0}^{Page} &=& *j_{D0}^{bs} - B\wedge *j_{D2}^{bs}\nn
&&
+ \frac{1}{2}B\wedge B\wedge
*j_{D4}^{bs} - \frac{1}{6}B\wedge B\wedge B\wedge *j_{D6}^{bs}
\label{d0d6}
\ena

Recall that for D6/D4/D2/D0 bound state, the brane sources
can be read off from the WZ term for D6-brane \p{wzd6}, the 
corresponding brane source charges are
\bea
Q_{D4}^{bs} &=& \int *j_{D4}^{bs} = \int (B+F)\nn
Q_{D2}^{bs} &=& \int *j_{D2}^{bs} = \int \frac{1}{2}(B+F)\wedge(B+F)\nn
Q_{D0}^{bs} &=& \int *j_{D0}^{bs} = \int \frac{1}{6}(B+F)\wedge(B+F)
\label{d6bs}
\wedge(B+F)\nn  
\ena
Putting Eqs.\p{d4d6}, \p{d2d6}, \p{d0d6} and \p{d6bs} together, we obtain
for D4-, D2-branes 
\EQ
Q_{D4}^{Page} = \int *j_{D4}^{Page} = \int F
\label{pd4d6}
\EN
\bea
Q_{D2}^{Page} &=& \int \Big(*j_{D2}^{bs} - B\wedge *j_{D4}^{bs} + 
\frac{1}{2}B\wedge B\wedge *j_{D6}^{bs}\Big)\nn
&=& \int \Big\{\frac{1}{2}(B+F)\wedge (B+F) - B\wedge (B+F) + \frac{1}{2}
B\wedge B\Big\}\nn
&=& \int \frac{1}{2}F\wedge F
\label{pd2d6}
\ena
and for D0-branes
\bea
Q_{D0}^{Page} &=& \int \Big(*j_{D0}^{bs} - B\wedge *j_{D2}^{bs}
+ \frac{1}{2}B\wedge B\wedge *j_{D4}^{bs}
- \frac{1}{6}B\wedge B\wedge B\wedge *j_{D6}^{bs}\Big)\nn
&=& \int \Big\{\frac{1}{6}(B+F)\wedge (B+F)\wedge (B+F)
- \frac{1}{2}B\wedge (B+F)\wedge (B+F)\nn
&& 
+ \frac{1}{2}B\wedge B\wedge (B+F) - \frac{1}{6}B\wedge B\wedge B\Big\}\nn
&=& \int\frac{1}{6}F\wedge F\wedge F
\label{pd0d6}
\ena
Eq.\p{pd0d6} shows that the terms $B\wedge B\wedge B$, $B\wedge B\wedge F$,
$B\wedge F\wedge F$ are cancelled with each other in the RR Page charge
of D0-branes living in D6-brane, which was expected in~\cite{wt},~\cite{amm}
and ~\cite{dm}.

\vs{5}
{\Large\bf 4. Summary and Discussion}
\vs{5}

So far we have developed new approach to consistently construct 
B-independent RR charges 
of D2p-branes in B fields.
As D-branes are sources of RR gauge fields, we have induced the equations
of motion and the nonvanishing (modified) Bianchi identities from
total action for IIA supergravity plus D-brane sources,
with which we have defined the duals of brane source currents for D2p-branes.
By making use of the relation $dA_{p}\wedge B_{q} = d(A_{p}\wedge B_{q}) - 
(-1)^{p}A_{p}\wedge dB_{q}$, we have plugged the equations for the
duals of brane source currents for D(2p+2n)-branes into
that for the D2p-brane iteratively, and found that the resulting
equations can be recast into that the left sides of equations
are exterior derivative and the right sides are localized objects.
This particular feature indicates that the right side localized objects can
be identified
as Page charges because of their conservation and locality,
which shows that the Page charges for D2p-branes can
be consistently defined from the brane source charges. Inserting
the brane source charges into the expression of the Page charges,
we have shown that all the Page charges are independent of the background
B fields which provides further evidence that
the RR Page charge should be quantized. In our explicit construction, 
it is highly nontrivial
that the B-dependent terms like $B\wedge B\wedge B$,
$B\wedge B\wedge F$, $B\wedge F\wedge F$ from different sources
are exactly cancelled with each other. Thus we conclude that
the properly defined RR charges of D-branes should be B-independent,
and we can identify them as Page charges which are conserved and
localized \footnote{When B fields are constant, all the brane
source charges are conserved, so in that case there is no
necessity to construct RR Page charge.}.

The IIA supergravity only contains 2-form $F^{(2)}$ and 4-form $F^{(4)}$  
field strength, until now we only considered D2-, D4-, D6-brane in B fields.
It would be interesting to study D8-brane in B fields and construct
the corresponding RR Page charges for D0-, D2-, D4-, D6-branes which couple
to D8-brane via WZ term. In this case, we should turn to massive type IIA
supergravity. In the above construction, we have assumed $H = dB$, 
however when it does not
hold in whole region, for instance, in the near-horizon region of
the NS5-brane background where the geometry is $R^{1,5}\times R_{\phi}
\times S^{3}$, we cover $S^{3}$ with two open sets 
$U = \{u_{\a}\}$, the RR Page charge defined on each open set with
$B_{\a}$ is conserved. It is interesting to see how two RR Page charges
defined in each open set are connected. It is probably related to brane
creation phenomenom~\cite{kz}. We hope to discuss the above questions in near
future.

\section*{Acknowledgement} 

I would like to thank B. Harms, M. Kreuzer, D. Marolf, H.J. Schnitzer
and P.K. Townsend for valuable discussion. This work is supported
by the Austrian Research Funds FWF under grant No. M597-TPH.

\newcommand{\NP}[1]{Nucl.\ Phys.\ {\bf #1}}
\newcommand{\AP}[1]{Ann.\ Phys.\ {\bf #1}}
\newcommand{\PL}[1]{Phys.\ Lett.\ {\bf #1}}
\newcommand{\CQG}[1]{Class. Quant. Gravity {\bf #1}}
\newcommand{\CMP}[1]{Comm.\ Math.\ Phys.\ {\bf #1}}
\newcommand{\PR}[1]{Phys.\ Rev.\ {\bf #1}}
\newcommand{\PRL}[1]{Phys.\ Rev.\ Lett.\ {\bf #1}}
\newcommand{\PRE}[1]{Phys.\ Rep.\ {\bf #1}}
\newcommand{\PTP}[1]{Prog.\ Theor.\ Phys.\ {\bf #1}}
\newcommand{\PTPS}[1]{Prog.\ Theor.\ Phys.\ Suppl.\ {\bf #1}}
\newcommand{\MPL}[1]{Mod.\ Phys.\ Lett.\ {\bf #1}}
\newcommand{\IJMP}[1]{Int.\ Jour.\ Mod.\ Phys.\ {\bf #1}}
\newcommand{\JHEP}[1]{J.\ High\ Energy\ Phys.\ {\bf #1}}
\newcommand{\JP}[1]{Jour.\ Phys.\ {\bf #1}}

\end{document}